\documentclass[table,twocolumn]{aastex62}
\usepackage{amstext}
\usepackage{amsmath}
\usepackage{apjfonts}
\usepackage[capbesideposition={top}]{floatrow}
\usepackage{float}
\usepackage{cancel}

\DeclareSymbolFont{cmletters}{OML}{cmm}{m}{it}
\DeclareMathSymbol{v}{\mathalpha}{cmletters}{"76}

\newcommand{\secref}[1]{\S\ref{#1}}

\newcommand{\beqar}{\begin{eqnarray}}
\newcommand{\eeqar}{\end{eqnarray}}

\newcommand{\machinf}{\mathcal{M}_\infty}
\newcommand{\ra}{R_{\rm a}}
\newcommand{\rg}{r_{\rm g}}

\newcommand{\cinf}{c_\infty}
\newcommand{\vinf}{v_\infty}
\newcommand{\rhoinf}{\rho_\infty}
\newcommand{\betainf}{\beta_\infty}

\newcommand{\beq}{\begin{equation}}
\newcommand{\eeq}{\end{equation}}

\definecolor{nick}{HTML}{006400}

\begin{document}
\title{Jet Formation in 3D GRMHD Simulations of Bondi-Hoyle-Lyttleton Accretion}

\correspondingauthor{Nicholas Kaaz}
\email{nkaaz@u.northwestern.edu}

\author[0000-0002-5375-8232]{Nicholas Kaaz}
\affiliation{Department of Physics \& Astronomy, Northwestern University, Evanston, IL 60202, USA}
\affiliation{Center for Interdisciplinary Exploration \& Research in Astrophysics (CIERA), Evanston, IL 60202, USA}

\author[0000-0003-2333-6116]{Ariadna~Murguia-Berthier}
\altaffiliation{NASA Einstein Fellow}
\affiliation{Center for Interdisciplinary Exploration \& Research in Astrophysics (CIERA), Evanston, IL 60202, USA}

\author[0000-0002-2825-3590]{Koushik Chatterjee}
\affiliation{Black Hole Initiative at Harvard University, 20 Garden Street, Cambridge, MA 02138, USA}

\author{Matthew T.P. Liska}
\affiliation{Institute for Theory and Computation, Harvard University, 60 Garden Street, Cambridge, MA 02138, USA
}

\author[0000-0002-9182-2047]{Alexander Tchekhovskoy}
\affiliation{Department of Physics \& Astronomy, Northwestern University, Evanston, IL 60202, USA}
\affiliation{Center for Interdisciplinary Exploration \& Research in Astrophysics (CIERA), Evanston, IL 60202, USA}

\begin{abstract} 
A black hole (BH) travelling through a uniform, gaseous medium is described by Bondi-Hoyle-Lyttleton (BHL) accretion. If the medium is magnetized, then the black hole can produce relativistic outflows. We performed the first 3D, general-relativistic magnetohydrodynamics simulations of BHL accretion onto rapidly rotating black holes using the code \verb|H-AMR|, where we mainly varied the strength of a background magnetic field that threads the medium. We found that the ensuing accretion continuously drags to the BH the magnetic flux, which accumulates near the event horizon until it becomes dynamically important. Depending on the strength of the background magnetic field, the BHs can sometimes launch relativistic jets with high enough power to drill out of the inner accretion flow, become bent by the headwind, and escape to large distances. While for stronger background magnetic fields the jets are continuously powered, at weaker field strengths they are intermittent, turning on and off depending on the fluctuating gas and magnetic flux distributions near the event horizon. We find that our jets reach extremely high efficiencies of $\sim100-300\%$, even in the absence of an accretion disk.
We also calculated the drag forces exerted by the gas onto to the BH, finding that the presence of magnetic fields causes drag forces to be much less efficient than in unmagnetized BHL accretion, and sometimes become negative, accelerating the BH rather than slowing it down. Our results extend classical BHL accretion to rotating BHs moving through magnetized media and demonstrate that accretion and drag are significantly altered in this environment.
\end{abstract}


\section{Introduction}
A black hole (BH) traversing through a gaseous medium is common in a wide variety of astrophysical scenarios, including common envelope evolution \citep{macleod_2015,macleod_2017,ari_2017,rosa_2020}, wind accretion in binary systems \citep{elmellah_2015,elmellah_2018}, and the passage of stellar-mass black holes through the disks of active galactic nuclei \citep{stone_2017,kaaz_2021}. As the BH travels through the medium, it will accrete mass and potentially produce energetic outflows that alter its environment. However, outflows are often magnetically-driven, yet the traditional description of accretion onto a moving source, ``Bondi-Hoyle-Lyttleton'' (BHL) accretion, is purely hydrodynamic \citep{hoyle_lyttleton_1939,bondi_1952,edgar_2004}. In reality, astrophysical plasmas are usually at least weakly magnetized. While magnetized BHL accretion has been studied generally \citep{lee_2014}, it is particularly intriguing if the central object is a BH, because then relativistic jets may be launched. Non-magnetized relativistic BHL accretion has been simulated in the past \citep{font_1998,penner_2013,lora-clavijo_2013,koyuncu_2014,gracia-linares_2015,lora-clavijo_2015,blakely_2015,cruz-osorio_2017}, as has 2D magnetized relativistic BHL accretion \citep{penner_2011}, but to the best of our knowledge, our work is the first to simulate 3D general-relativistic magnetohydrodynamic (GRMHD) BHL accretion. 

A widely accepted way of powering relativistic jets is the Blandford-Znajek (BZ) mechanism \citep{blandford_znajek_1977}. This mechanism converts rotational energy from the BH into collimated outflows that propagate to large scales. The BZ mechanism is activated if there is a large enough accumulation of magnetic flux onto the rotating BH. In BHL accretion, a steady supply of gas is provided to the BH. However, the magnetic fields threading the gas cannot cross the event horizon. Instead, the magnetic threads become anchored within the inner accretion flow and gradually build up.  Hypothetically, this build up could continue until the flow becomes `magnetically arrested', which is when the Lorentz force begins competing with gravity \citep{narayan_2003}. This suggests that even if a BH travels through a weakly-magnetized medium, it still might accumulate enough magnetic flux to power strong BZ jets after enough time has elapsed. 


If a BH produces relativistic jets, it will deposit energy and momentum into its surroundings and impede the accretion flow. This can potentially alter the gravitational drag exerted on the BH by the gas supply. Recent works have found that depending on the nature of the feedback in BHL accretion, the gas may exert \textit{negative} drag, accelerating the traveling BH rather than slowing it down \citep{li_2020,gruzinov_2020}.  Even if the BH is non-rotating and thus unable to launch a jet, dynamically-important magnetic fields would likely still affect the accretion flow enough to modify drag. This could have particularly important consequences for common envelope evolution, where the efficiency of drag partially dictates the final configuration of the binary. 

In the rest frame of the traveling BH, the medium behaves as a supersonic, gaseous wind. As relativistic jets traverse this wind, they may become bent and possibly disrupted, depending on the jet power relative to the wind thrust. Bent double radio sources are a subclass of AGN jets that exhibit a bent morphology, which occur due to the interaction of the jet with the intergalactic medium \citep{begelman_1979,freeland_wilcots_2011,morsony_2013,musoke_2020}, although this is somewhat different from our scenario as the winds that bend these jets do not supply the accretion flow. Jet bending may also occur in X-ray binaries (XRBs), either because the BH is moving at supersonic speeds through the interstellar medium (ISM) due to its natal supernova kick \citep{heinz_2008,wiersema_2009,yoon_2011}, or because the stellar wind from a binary companion applies ram pressure to the jet \citep{yoon_2015}.

If a binary system undergoes unstable Roche lobe overflow, it can lead to common envelope evolution. Depending on the mass ratio, the dynamics within the common envelope can be modeled with a BHL-like wind tunnel \citep{macleod_2015,macleod_2017,rosa_2020}. In the case that the embedded companion is a BH, then feedback produced by the BH can be deposited into the envelope of the host star, potentially unbinding it and producing novel electromagnetic transients \citep{moreno-mendez_2017,lopez-camara_2020}. If the envelope unbinds before the BH merges with the core of the host star, then this will alter the configuration of the binary remnant, which could hinder the viability of common envelopes as formation channels for LIGO/Virgo sources \citep[e.g., ][]{olejak_2021}.

In this work, we simulate magnetized BHL accretion in three dimensions onto spinning BHs with aims to characterize the resulting accretion flow and understand the consequences of the feedback processes. In \secref{sec:approach}, we review the characteristic scales of BHL accretion, lay out our expectations for jet formation, and describe the simulations we perform. In \secref{sec:results}, we present the results of our simulations. In \secref{sec:discussion}, we discuss the consequences of our results for various astrophysical environments and summarize our main findings. 


\section{Approach}
\label{sec:approach}

\subsection{Characteristic scales}
\label{sec:approach:scales}

In classical BHL accretion, a gaseous wind engulfs a massive object, which gravitationally focuses the wind in a downstream wake \citep[for a review, see ][]{edgar_2004}. The object, which has mass $M$, accretes directly from this wake. The accretion radius, $\ra$, defines the range of the BH's influence on the wind, within which all streamlines are accreted,

\begin{equation}
    \ra = \frac{2GM}{\vinf^2} = \left(\frac{\vinf}{c}\right)^{-2}2\rg,
    \label{eq:accretion_radius}
\end{equation}

where $\vinf$ is the velocity of the wind and $\rg=GM/c^2$ is the BH gravitational radius. We note that some authors use alternative conventions where they define the accretion radius using the sound speed of the wind, $\cinf$, as well \citep[$\ra\equiv GM/(\cinf^2+\vinf^2)$, e.g. ][]{cruzosorio_2020}, which can be more appropriate at lower Mach numbers. Assuming that all streamlines propagate ballistically, the characteristic accretion rate of the system can be estimated by multiplying the mass flux of the wind, $\rhoinf\vinf$, by the geometric cross-section of the accretion radius, $\pi \ra^2$,

\begin{equation}
    \dot{M}_{\rm HL} = \pi\ra^2\rhoinf\vinf = \frac{4\pi\rg^2\rhoinf c}{(\vinf/c)^{3}}
    \label{eq:hl_accretion_rate}
\end{equation}

Here, $\rhoinf$ is the ambient gas density. $\dot{M}_{\rm HL}$ is the `Hoyle-Lyttleton' accretion rate \citep{hoyle_lyttleton_1939}, and is found to be correct to order-unity in numerical simulations \citep{blondin_2012}. The characteristic accretion timescale is

\begin{equation}
    \tau_{\rm a} = \ra/\vinf = \frac{2GM}{\vinf^3}
    \label{eq:accretion_timescale}
\end{equation}

The accretion flow typically reaches steady state at about $\sim10-20\,\tau_{\rm a}$. We note that while these are Newtonian definitions, they still hold in our analysis of general-relativistic flows, as the accretion radius is much larger than the event horizon.

In this work, we assume that the ambient medium is threaded with a uniform strength magnetic field oriented in a single cartesian direction. This adds two new free parameters: the ratio of gas to magnetic pressure in the ambient medium, $\betainf$, and the orientation of the background field, $\hat{B}$, where $\vec{B}_\infty = B_\infty \hat{B}$. In our setup, magnetic field lines are dragged inwards with the accretion flow, and, unless the magnetic flux efficiently escapes downstream with the wind, the field lines will accumulate near the BH. Once the forces due to the magnetic pressure and the BH gravity become comparable, the accretion flow becomes `magnetically arrested' (denoted MAD\footnote{The abbreviation `MAD' stands for magnetically arrested disk \citep{narayan_2003}. There is no angular momentum in our accretion flow and thus no disk, but we refer to it as a MAD anyways because this is the term that is commonly used in the literature.}) \citep{narayan_2003}, and the BH magnetic flux saturates. We can express this condition by equating the magnetic pressure with the gas pressure,

\begin{equation}
    \frac{B_{\rm MAD}^2}{8\pi} = P_{\rm g}
\end{equation}

If we assume that the gas is roughly virial, which is appropriate for a flow that doesn't cool, then the sound speed is $\approx \sqrt{GM/r}$. Assuming an ideal gas equation of state, we can then rewrite the previous expression as,

\begin{equation}
    \frac{B_{\rm MAD}^2}{8\pi} = \frac{GM\rho(r)}{\gamma r}
\end{equation}

Since we are interested in the strength of the magnetic field that launches the jets, we evaluate this expression at the event horizon radius, which we take to be roughly $\approx r_{\rm g}$ for a nearly maximally-rotating BH. By assuming that the density profile follows a Bondi-like profile, $\rho(r) = \rhoinf(r/R_{\rm a})^{-3/2}$ \citep[][]{shapiro_teukolsky_1983}, we find

\begin{equation}
    B_{\rm MAD} = \sqrt{4\pi\rhoinf\vinf^2/\gamma}(\rg/\ra)^{-5/4},
\end{equation}

suggesting the magnetic field required to create a magnetically arrested flow is only a function of the density and velocity of the incident wind. The term $\rg/\ra$ also only depends on $\vinf/c$ (Eq. \ref{eq:accretion_radius}), which we leave in its given form to highlight the importance of the size of $\ra$. Then, the critical magnetic flux is $\Phi_{\rm MAD}\sim B_{\rm MAD}\times({\rm Area}) = 2\pi \rg^2 B_{\rm MAD}$. If we assume that the magnetic field vector is orthogonal to the wind velocity vector, the magnetic flux transported by the wind within time $dt$ is given by, 

\begin{equation}
    d\Phi = 2\ra B_{\infty} \vinf dt
\end{equation}

If we integrate in time until $\Phi(t=\tau_{\rm MAD})=\Phi_{\rm MAD}$, we can obtain a characteristic timescale for the flow to become magnetically arrested,

\begin{equation}
    \tau_{\rm MAD} = \pi\frac{\ra}{\vinf}\machinf\sqrt{2\betainf\gamma}(r_{\rm g}/R_{\rm a})^{3/4},
    \label{eq:ma_timescale}
\end{equation}

which we can express as, 

\begin{equation}
    \tau_{\rm MAD} = 0.33 \left(\frac{\beta_\infty}{10}\right)^{1/2}\left(\frac{\machinf}{2.45}\right)\left(\frac{200\,\rg}{\ra}\right)^{3/4}\tau_{\rm a}
\end{equation}

This suggests that even when the background magnetic pressure is highly subdominant to the gas pressure, the flow can become magnetically arrested on the order of an accretion timescale. Realistically, this timescale is a lower limit, as it assumes all of the magnetic flux that is transported within the accretion cross-section can reach the event horizon. Whether or not the flow can develop into a MAD in practice depends on how efficiently the inner accretion flow can hold onto its magnetic flux, which we explore in our numerical simulations. 

\begin{figure*}
    \centering
    \includegraphics[width=\textwidth]{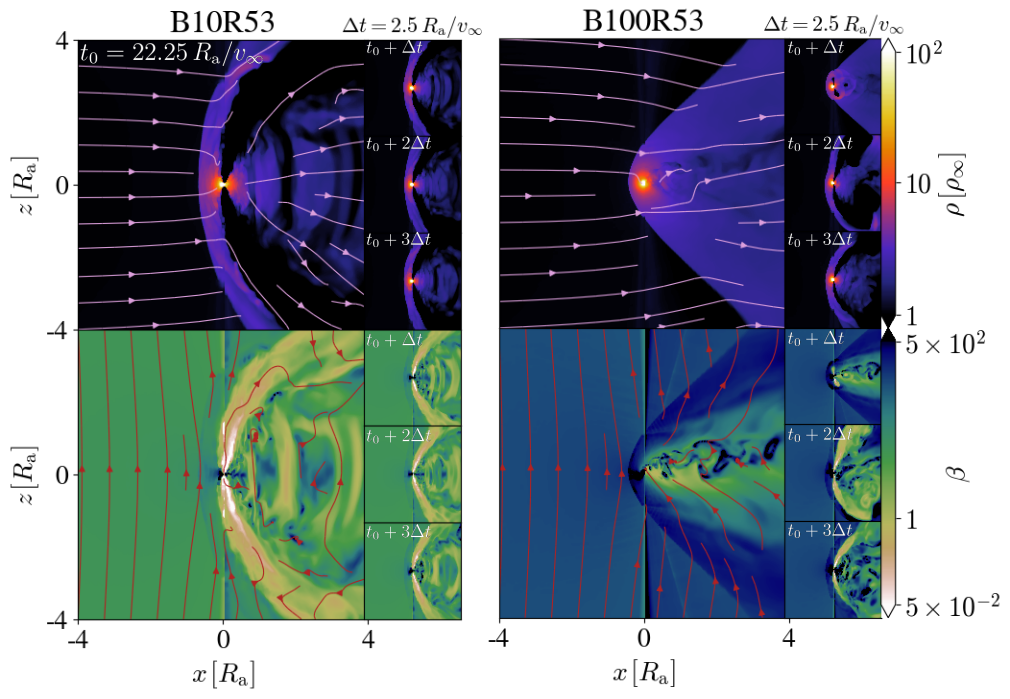}
    \caption{Black holes accreting from a supersonic, magnetized wind sometimes produce jets strong enough to escape their surrounding high-density medium, depending on the strength of the ambient magnetic field. We depict $x-z$ slices of the fluid frame gas density ($\rho$) and the ratio of gas-to-magnetic pressure ($\beta$) in our $\betainf=10$ (left) and $\betainf=100$ (right) simulations. The wind travels in the $+\hat{x}$ direction, and the background magnetic field and BH rotation axis are oriented along the $+\hat{z}$ direction. In the $\rho$ panels, velocity streamlines are shown in pink, and in the $\beta$ panels, magnetic field lines are shown in red. The large panels are each shown at a time $t_0\approx22.25\,\ra/\vinf$.  
    Each figure is accompanied by three-panel film strips, which depict the same simulation at later times, separated by $\Delta t=2.5\,\ra/\vinf$ intervals and highlight the diversity in the flow morphology. While at $\betainf=10$ the jet is dominant throughout the simulation, at $\betainf=100$ the jet turns on and off intermittently.}
    \label{fig:film_strips}
\end{figure*}

\subsection{Numerical Method and Simulation Setup}
\label{subsec:simulations}

\begin{table}[]
\begin{tabular}{llll}
 & $\beta_\infty$ & $a$ & $\gamma$ \\\hline
B1R53              & 1             & 0.9 & 5/3      \\
B10R53              & 10            & 0.9 & 5/3      \\
B50R53              & 50             & 0.9 & 5/3   
  \\
B100R53              & 100             & 0.9 & 5/3      \\
B200R53              & 200             & 0.9   & 5/3      \\
B10R43              & 10             & 0.9 & 4/3      \\
B10NR53              & 10             & 0.0 & 5/3      \\
NBNR53              & $\infty$             & 0.0 & 5/3      \\
\end{tabular}
\caption{Parameters used in all simulations are listed here. The naming scheme is: `B' followed by $\betainf$ or `NB' for a non-magnetized medium, `R' or `NR' for rotating ($a=0.9$) or non-rotating BHs, `53' or `43' for $\gamma=5/3$ or $4/3$. See \secref{subsec:simulations} for a description of each parameter.}
\label{table:sims}
\end{table}

We evolve our simulations using the GPU-accelerated, three-dimensional general-relativistic magnetohydrodynamic code \verb|H-AMR| \citep{HAMR}. \verb|H-AMR| uses a spherical polar grid that is uniform in ${\rm log}r$ and centered on the BH. We use a Kerr metric where the BH spin vector is in the $+\hat{z}$ direction. We also use multiple numerical speed-ups that make \verb|H-AMR| well-suited to problems with large scale separation, such as the one studied here. This includes five levels of  local adaptive time-stepping and $\theta$-dependent static refinement criteria \citep[for details see][]{HAMR}. The base resolution of our grid is $({\rm N}_{r}\times{\rm N}_\theta\times{\rm N}_\phi)=(768\times192\times256)$. We use horizon-penetrating Kerr-Schild coordinates and place the inner radial boundary sufficiently deep within the BH event horizon such that the boundary is causally disconnected from the BH exterior. By placing the outer radial boundary at a sufficiently large radius, $R_{\rm out} = 2\times10^5 r_{\rm g}$, we ensure that it is causally disconnected from the accretion flow. Both radial boundaries use inflow boundary conditions. The polar boundary condition is transmissive and the azimuthal boundary condition is periodic \citep{liska_2018}. 

We evolve our simulations in a scale-free fashion such that $G=M=c=\rhoinf=1$, with a maximum runtime of $t=50\,\ra/\vinf=10^5\,r_{\rm g}/c$. We set the non-relativistic Mach number of the wind to $\machinf=2.45$, the accretion radius to $\ra=200\rg$ and the magnetic field direction to $\hat{B}=\hat{z}$ in all simulations. We choose our value of $\machinf$ because a mildly supersonic wind is more likely than a highly supersonic wind in most realistic astrophysical applications. Our choice of $\machinf$ and $\ra$ also sets the sound speed, which is $c_\infty \approx 0.04$. We choose $\ra/\rg$ to be as large as computationally feasible: realistic values are $\ra\sim10^4-10^6\,\rg$, but this would take too long to simulate. Even then, in full GRMHD our simulated ratio of $\ra/\rg$ is as large as the highest resolution Newtonian hydrodynamic studies \citep[where $\rg$ is replaced by the `sink radius', ][]{xu_2019} and is substantially larger than in relativistic hydrodynamic studies, which reach about $\ra/\rg\sim17$ in 3D \citep{blakely_2015}. We present eight simulations where we primarily vary the value of the ambient gas-to-magnetic pressure ratio, $\beta_\infty$. We use an ideal gas equation of state, where we typically explore an adiabatic index $\gamma=5/3$, but in one simulation test $\gamma=4/3$. In most of our simulations, we use a dimensionless BH spin of $a=0.9$, but also consider $a=0$ in one of our simulations. We list our specific parameter choices for all simulations in Table \ref{table:sims}. Additionally, in Appendix \ref{app:bhl}, we present an unmagnetized BHL run that we compare with previous hydrodynamic results. 

\section{Results}
\label{sec:results}
In the following sections we describe the outcome of our simulations.

\subsection{Flow morphology}
\label{sec:results:flow}

\begin{figure*}
    \centering
    \includegraphics[width=\textwidth]{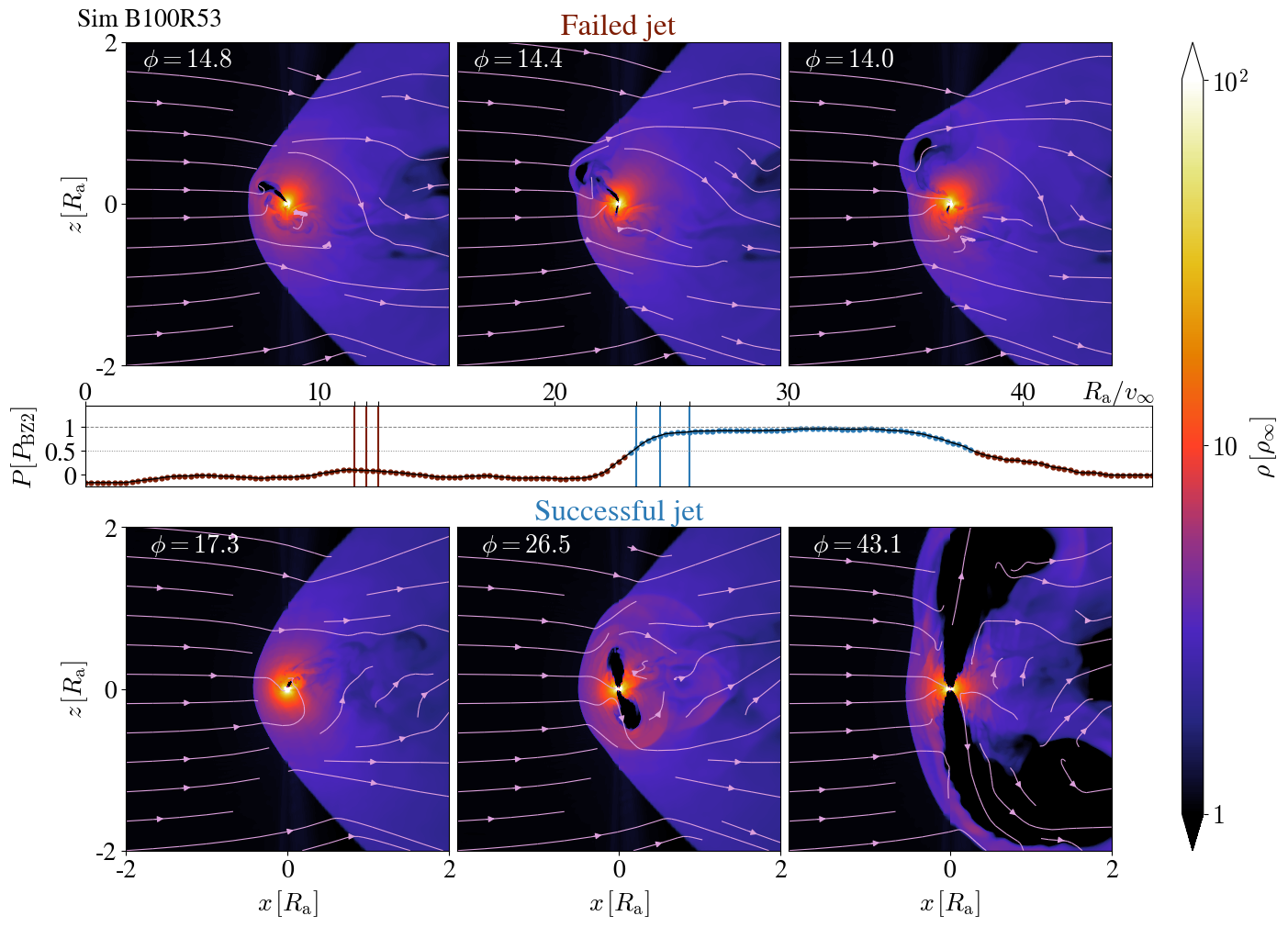}
    \caption{Threaded with vertical magnetic fields, our BHs attempt to launch jets, which either succeed or fail depending on how much magnetic flux they have accumulated. Here, we illustrate the attempts at jet launching in simulation B100R53. In the top row, we depict consecutive snapshots in time of the BH launching a nascent jet that is quickly quenched. In the bottom row, we depict another time series where the jet is instead successful. Each snapshot depicts gas density, overlaid with velocity streamlines in pink. The success or failure of the jets depends on the BH dimensionless magnetic flux, $\phi$, which is given in the top-left corner of each panel. In the middle row, we plot the jet luminosity as a function of time. Blue dots indicate a successful jet, whereas red dots indicate a failed jet. We define `successful' as $P>0.5\,P_{\rm BZ2}$; while this is a rough criterion, we have found that it effectively picks out times when a jet launches, as seen in this plot.}
    \label{fig:jetStruggle}
\end{figure*}

In our simulations, we mainly vary the initial plasma beta parameter, $\betainf$, which spans the full range from unmagnetized to strongly magnetized media. We illustrate the effect of changing $\betainf$ in Figure \ref{fig:film_strips}. This figure shows the fluid frame gas density $\rho$ and plasma $\beta$ in $x-z$ slices of the flow for simulations B10R53 and B100R53, which differ only in their value of $\betainf$. Let us first turn to simulation B10R53 in the left column. Here, the background magnetic fields are strong enough ($\betainf=10$) to power the jet throughout the entire simulation runtime. These jets appear as the low-density (black) polar regions in the $\rho$ snapshot and the magnetically-dominated (white-green) polar regions in the $\beta$ snapshot. We can see that near the BH ($\lesssim\ra$), the jets follow the $\pm\hat{z}$ axes, but are quickly bent by the ram pressure of the incident wind. The bending angle quickly increases with increasing distance away from the BH, and asymptotically approaches $90^\circ$ (not depicted). The ram pressure of the jets as they slam into the accretion flow also causes the bow shock to expand, leaving a thin high-density layer between the bow shock and the jets. This will become more apparent when we examine the time series in Figure \ref{fig:jetStruggle} later on. 

Next, let us turn to simulation B100R53 in the right column of Fig. \ref{fig:film_strips}. Here, the background magnetic fields are a factor of 10 weaker and not strong enough to produce persistent jets. In the main panels, we see the same features that are characteristic of unmagnetized BHL accretion: the supersonic wind, travelling in the $+\hat{x}$ direction, engulfs the BH. As the velocity streamlines (pink, in the density snapshot) focus toward the BH, they form a bow shock. Streamlines within $\sim\ra$ of the BH are accreted, and in the classic problem this accretion occurs in an over-dense downstream wake. However, the accumulation of magnetic flux disrupts the accretion flow in the downstream region. This is most apparent in the $\beta$ snapshot, where we see alternating regions of highly magnetized ($\beta\lesssim1$) and weakly magnetized ($\beta\gtrsim10^2$) plasma downstream of the BH. The magnetic fields bend radially in this region and tangle up (red streamlines in $\beta$ snapshot). We pair each panel with a film strip of three snapshots separated by $2.5\,\ra/\vinf$ time intervals. In the case of the B100R53 simulation, we see that the jets launch later on. We hypothesize that this occurs once a threshold amount of magnetic flux threads the black hole horizon, which we explore further in \secref{sec:results:jetlaunching}. At even later times, the jet will turn off again, suggesting that there is a duty cycle like behavior for the jets that depends on how efficiently the magnetic flux can escape the BH.

\subsection{Launching the jets}
\label{sec:results:jetlaunching}

In \secref{sec:results:flow}, we learned that depending on the value of $\betainf$, jets can be launched continuously or sporadically\footnote{It is possible that there is a high enough value of $\betainf$ above which jets never launch, but we are unable to test this in the present work.}. In this section, we will try and get a better sense of what allows the jets to launch in our simulations. We will assume that the jets form through a Blandford-Znajek (BZ) mechanism \citep{blandford_znajek_1977}. In this paradigm, jets launch when a rotating BH is threaded with large-scale poloidal magnetic flux. The jets form by the conversion of the rotational energy of the BH into Poynting flux that the jets transport outward. We adopt the following expression for the spindown power of the BH, which is accurate when $a\lesssim0.95$ \citep{tchekhovskoy_2011},
\begin{equation}
    P_{\rm BZ2}= \frac{c}{96\pi^2r_{\rm g}^2}\Phi^2\left(\frac{a}{1+\sqrt{1-a^2}}\right)^2
    \label{eq:lbz2}
\end{equation}
Here,\footnote{The expression for $P_{\rm BZ2}$ in Equation \ref{eq:lbz2} is given in Gaussian units. We use the subscript `2' because this expression is a second-order expansion of the BH spin-down luminosity in powers of the dimensionless angular frequency of the BH, $\omega_{\rm H}$, where $\omega_{\rm H} \equiv a/(1+\sqrt{1-a^2})$.} $\Phi$ is the magnetic flux threading the event horizon of the BH. For a jet to be successfully launched, it needs enough power to overcome the hydrodynamic pressure of the surrounding accretion flow. For a fixed $a$, Equation \ref{eq:lbz2} is only a function of $\Phi$, suggesting that the success or failure of the jet mainly depends on how efficiently magnetic flux can be accumulated onto the BH. 

In Figure \ref{fig:jetStruggle}, we examine two separate periods of time for simulation B100R53. In the first, a nascent jet is launched, but is quickly expunged by the gas (top row, labeled `Failed jet'). In the second, a jet is launched, and then rapidly grows in power, clearing out the gas in the polar regions (bottom row, labeled 'Successful jet'). In each panel, we write the dimensionless magnetic flux ($\phi\equiv\Phi/\sqrt{\dot{M}}$) in the top left corner, in Gaussian units. We see that in the case of the failed jet, the highest value of $\phi$ is $14.4$, while for the successful jet, $\phi$ begins at $17.3$. While the precise `critical` value of $\phi$ to launch jets is unclear and depends on the distribution of gas near the BH, this suggests that it is in the range of $\sim15-17$. 

Between the bottom and top rows of Fig. \ref{fig:jetStruggle}, we plot the jet power, $P$, normalized to the measured value of $P_{\rm BZ2}$ as a function of time. Here, we calculate the jet power as,
\begin{equation}
    P=\dot{M}-\dot{E}
    \label{eq:Ljet},
\end{equation}
where the mass accretion is
\begin{equation}
    \dot{M} = -\int \rho u^r \sqrt{-g}d\theta d\varphi
    \label{eq:mdot}
\end{equation}
and the total energy flux (including the rest-mass energy flux) is,
\begin{equation}
    \dot{E} = \int T^{r}_{t} \sqrt{-g}d\theta d\varphi,
    \label{eq:edot}
\end{equation}
where $u^r$ is the radial component of the contravariant four-velocity in Kerr-Schild coordinates and $T^{r}_{t}$ is the component of the mixed stress-energy tensor associated with the radial transport of energy flux. The expressions for $\dot{M}$ and $\dot{E}$ use the sign convention so that positive values indicate that mass and energy are entering the BH. To calculate $P_{\rm BZ2}$, we use the following expression for magnetic flux,
\begin{equation}
    \Phi = \frac{1}{2}\int|B^r|\sqrt{-g}d\theta d\varphi,
    \label{eq:magnetic_flux}
\end{equation}
where $B^r$ is the radial contravariant component of the three-vector magnetic field. We measure $\Phi$ at the event horizon but compute $\dot{M}$ and $\dot{E}$ at $\simeq5\,r_{\rm g}$. This is because at smaller radii, the jet sometimes reaches low enough densities that the density floor in the code is triggered, which artificially changes density-dependent quantities. Finally, we define a jet to be `successful' when $P\geq0.5P_{\rm BZ2}$\footnote{We acknowledge that this is a somewhat arbitrary criterion. The boundary between a jet being `on' and `off' is ambiguous, so we chose to use a simple criterion that qualitatively captures the presence of a jet.}. Then, in the middle panel of Fig. \ref{fig:jetStruggle}, the blue dots indicate a successful jet and the red dots indicate a failed jet. Each of the two rows of density snapshots is correspondingly marked in the time series with a trio of red vertical lines for the failed jet and a trio of blue vertical lines for the successful jet. We can immediately see that our jet success criterion tracks the morphological evolution of the jet. For the failed jet, $P$ remains much lower than $P_{\rm BZ2}$. However, when the jet is successfully launched, $P/P_{\rm BZ2}$ rapidly rises before plateauing to an equilibrium value of $\sim1$, as expected from theory. 

\begin{figure}
    \centering
    \includegraphics[width=\textwidth]{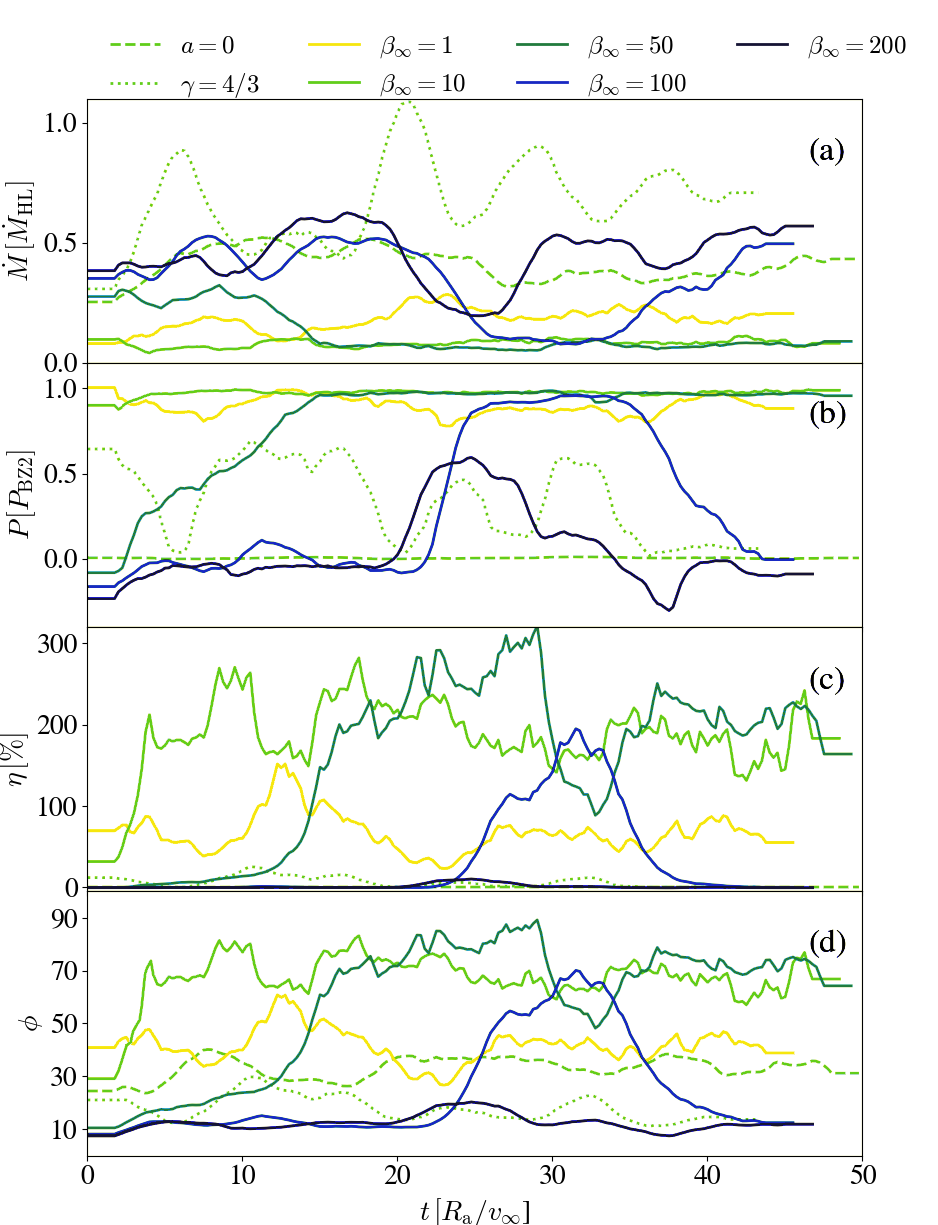}
    \caption{When jets are active, the feedback they exert on the infalling gas suppresses $\dot{M}$. This can be seen in the time series of various quantities, as depicted here. From top to bottom, the panels shown are: $\dot{M}$, the mass accretion rate; $P$, the jet luminosity (in units of $P_{\rm BZ2}$, see Eq. \ref{eq:lbz2}); $\eta$, the jet efficiency; and $\phi$, the dimensionless magnetic flux. Each quantity is calculated using smoothed profiles of $\dot{M}$, $\dot{E}$ and $\Phi$ that are averaged over a moving window of duration $\sim3.75\,\ra/\vinf$. This is done to make the plot more readable, and in Fig. \ref{fig:app_timeSeries}, we show the same figure without the averaging. In some of the $\betainf=10$ curves in Fig. \ref{fig:timeSeries}, a portion of the data at early times is omitted. This is because the mass accretion rate very briefly turns negative, making some calculated quantities undefined. Because we smooth the data presented here, this undefined calculation propagates to nearby values as well.}
    \label{fig:timeSeries}
\end{figure}



We can see that the lessons drawn from Fig. \ref{fig:jetStruggle} are broadly consistent with our other simulations by studying the curves in Figure \ref{fig:timeSeries}, where we plot the time dependence of several quantities for each simulation. In panel (a), we plot the mass accretion rate $\dot{M}$, which is generally time-dependent and can dip to values far below the analytic rate $\dot{M}_{\rm HL}$, particularly in strongly-magnetized media such as $1<\betainf<50$. The reason for this is elucidated in panel (b), where we plot the jet power $P$ in units of the Blandford-Znajek jet power, $P_{\rm BZ2}$. Here, we can see that whenever $\dot{M}$ drops in panel (b), this is accompanied by a rise in the jet power, indicating that feedback due to the jet is ejecting part of the gas supply. We also see that whenever a jet is launched, $P/P_{\rm BZ2}$ saturates to values near unity, as theoretically expected. Exceptions to this include our $\gamma=4/3$ and $\betainf=200$ simulations, where the maximum value of $P/P_{\rm BZ2}$ is closer to $0.5$. We hypothesize that this is because these jets are more intermittent and unable to reach their equilibrium value before turning off. When $\betainf=200$, this is because there is less background magnetic flux available. When $\gamma=4/3$, this is because the flow is much more compressible and reaches higher densities, lowering the dimensionless magnetic flux threading the BH.  

In panel (c), we depict the jet efficiency $\eta$, defined as $P/\dot{M}\times100\%$. We can immediately see that $\eta$ reaches extremely large values of $200-300\%$, especially when $\betainf\sim10-50$. This is in the upper range of values measured in simulations of magnetically-arrested disks \citep{tchekhovskoy_2011}, suggesting our flow here is also magnetically arrested. This is further supported in panel (d), where we depict the dimensionless magnetic flux $\phi$. Here, we see that our large jet efficiencies are also associated with values of $\phi$ that are as high as $\sim50-80$. This is consistent with our arguments given in \secref{sec:approach:scales}, where we suggested that despite having relatively weak background magnetic fields, the persistent accumulation of magnetic flux within the accretion cross-section will cause the flow to become magnetically-arrested within a few accretion timescales (e.g., Eq. \ref{eq:ma_timescale}). Interestingly, this is \textit{not} true when $\gamma=4/3$. Since the gas is much more compressible at lower adiabatic indices, the mass accretion rate is higher, but the dimensionless magnetic flux is only around $\sim10$. This seems to suggest more compressible flows are worse at holding onto their magnetic flux. This could have implications for BHL-like flows where $\gamma=4/3$ or lower adiabatic indices are appropriate, such as radiation pressure dominated flows (i.e., BH common envelopes, as discussed in \secref{sec:discussion}) or flows where cooling is efficient.
\subsection{Scalings with $\betainf$}
\begin{figure}
    \centering
    \includegraphics[width=\textwidth]{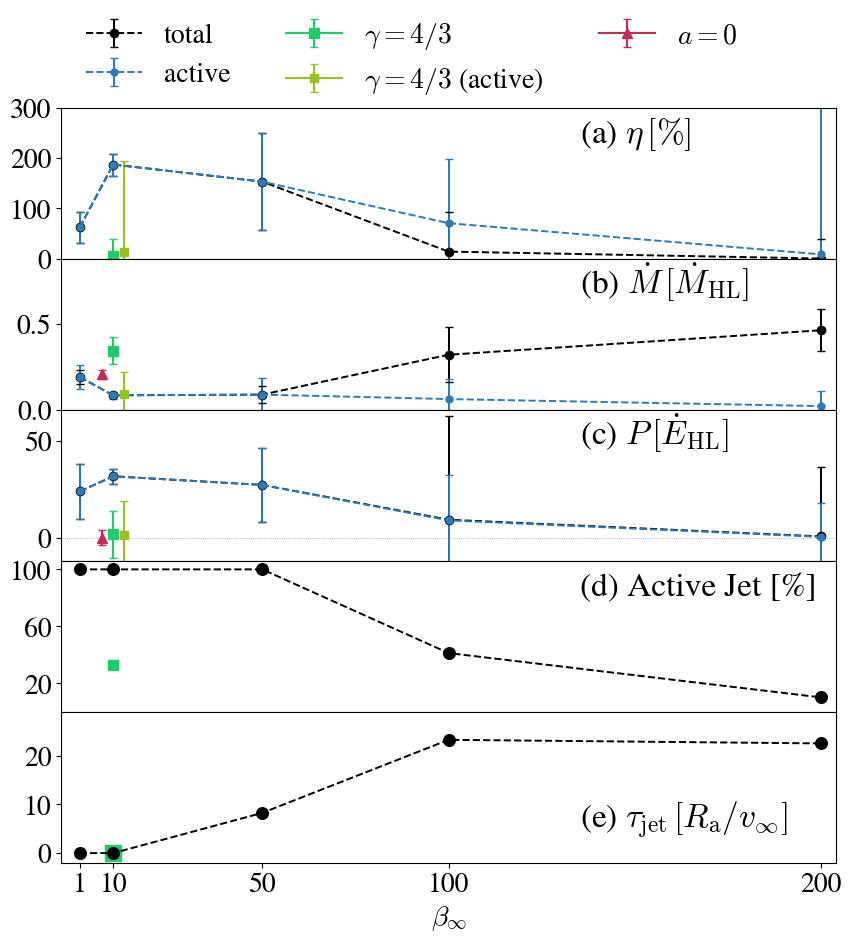}
    \caption{The strength and activity of our jets increase with decreasing $\betainf$, while the mass accretion rate decreases. We show this here by depicting various quantities as a function of $\betainf$ for each simulation. The time-variable quantities (panels a-c) are time-averaged at times $>10\,\ra/\vinf$. From top to bottom, the panels shown are: (a) $\eta$, the jet efficiency; (b) $\dot{M}$, mass accretion rate; (c) $L$, the jet luminosity; (d) the percentage of the time the jet is active for (where 'active'` is defined as $L>0.5L_{\rm BZ2}$); (e) the time until a jet is first launched.}
    \label{fig:betaSeries}
\end{figure}

In Figure \ref{fig:betaSeries}, we present various quantities as a function of $\betainf$. First, in panel (a), we show the time-averaged value of the jet efficiency\footnote{We calculate $\eta$ using the time-averaged values of $\dot{M}$ and $\dot{E}$, rather than time-averaging $\eta$ itself.} $\eta$ as a function of $\betainf$ for each simulation. We also depict the corresponding $\pm1$ standard deviation error bars. In general, we see a declining trend in $\eta$ with higher $\betainf$. This is expected, as background media with weaker magnetic fields will cause the BH to launch weaker jets. Simulation B1R53 is an exception to this trend; here, $\eta$ decreases despite the lower value of $\betainf$. This is because at $\betainf=1$, magnetic fields are dynamically important before accretion even begins. Perhaps counter-intuitively, this makes the jets less powerful. In all other simulations, the gas can readily drag magnetic flux with it to the event horizon. If magnetic fields are initially as strong as the gas, then the gas is less effective at dragging field lines with it. 

These trends are supported by panel (b) of Fig. \ref{fig:betaSeries}, where we plot the time-averaged mass accretion rate $\dot{M}$ and its $\pm1$ standard deviation error bars. Here, we see that $\dot{M}$ increases with increasing $\betainf$, and is anti-correlated with $\eta$ in panel (a). We also highlight the differences in the black and blue data points, where black points indicate a time-average over periods when the jet is both quiescent and active (labeled `total') and blue points indicate a time-average only over periods when the jet is active (labeled 'active'). We see that the increase of $\dot{M}$ with increasing $\betainf$ holds when averaging over all activity periods, but that $\dot{M}$ is relatively flat as a function of $\betainf$ if we only include periods when the jet is active. Yet, as seen from panel (a), the jet efficiency, even when measured only during active phases, is lower when $\betainf$ is higher. This suggests that the $\dot{M}-\betainf$ relation is modulated by how often the jet is powered, rather than how powerful the jet itself is. 

In panel (c) of Fig. \ref{fig:betaSeries}, we plot the time-averaged jet power $P$ along with its $\pm1$ standard deviation error bars in units of $\dot{E}_{\rm HL}\equiv\frac{1}{2}\dot{M}_{\rm HL}\vinf^2$, which is the energy flux that the BH captures from the wind. This is in contrast to Fig. \ref{fig:timeSeries}, where we plot $P$ in units of the Blandford-Znajek jet power. We see that the overall jet power decreases with increasing $\betainf$. $P$ also generally exceeds $\dot{E}_{\rm HL}$, especially when $\betainf$ is low, suggesting that the jets produce enough power to dominate the wind. Much of this energy will, however, be transported to scales $\gg\ra$ (e.g., Figs. \ref{fig:film_strips} and \ref{fig:jetStruggle}). In panel (d), we measure what percentage of time our jets are active for. For $1<\betainf<50$ and $\gamma=5/3$, the jets are active $100\%$ of the time. At weaker background magnetic field strengths, the jets become sporadic, and at $\betainf=200$ they are only on $10-15\%$ of the time. While the trend clearly suggests that higher $\betainf$ leads to jets being turned on less often, it remains unclear if a high enough $\betainf$ will prevent them from ever turning on. At $\gamma=4/3$, the jet is also sporadic even though $\betainf=10$, and is only active roughly $\sim30\%$ of the time. Finally, in panel (e), we measure the time delay until a jet is first turned on. It is nearly instantaneous when $\betainf=1$ or $10$, is slightly delayed when $\betainf=50$ ($\tau_{\rm jet}\sim10\,\ra/\vinf$), and is significantly delayed at $\betainf=100$ and $200$ ($\tau_{\rm jet}\sim25\,\ra/\vinf$). While $\tau_{\rm jet}$ does have a general trend with $\betainf$, we expect that the precise time at which the jet is launched is somewhat stochastic, as it depends on the highly variable distributions of gas and magnetic flux in the inner accretion flow. 

\begin{figure*}
    \centering
    \includegraphics[width=0.9\textwidth]{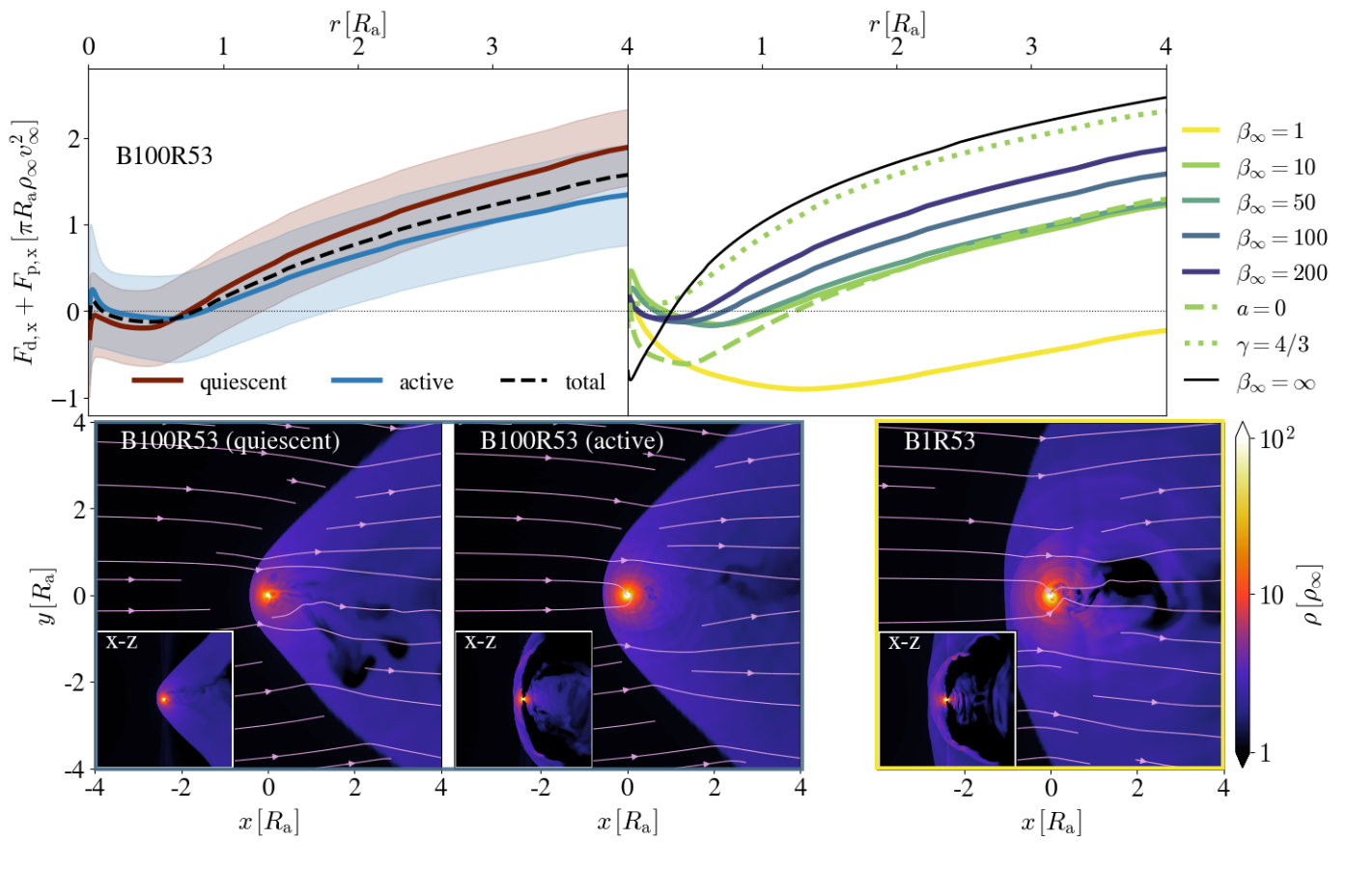}
    \caption{The drag forces exerted on the BH by the gas are less efficient for stronger background magnetic fields, and can become negative when the gas density near the BH is higher in the upstream region than in the downstream region. This effect is pronounced most strongly when $\betainf=1$, where drag is negative up to radii $\gtrsim4\,\ra$. \textbf{Top row.} Here, we plot radial profiles of the drag force exerted on the black hole. The drag forces depicted are the addition of the linear momentum that is accreted and the gravitational drag from the gas integrated over a volume of radius $r$. In the right panel, we plot the drag forces for each simulation, averaged over times $>10\,\ra/\vinf$. We also include the instantaneous drag force for a steady-state unmagnetized run (black). In the left panel, we time-average drag in simulation B100R53, split into times when the jet is active (blue) and quiescent (red), with corresponding $\pm1$ standard deviation shaded regions. \textbf{Bottom row.} Here, we plot three slices of gas density in the $x-y$ plane, with corresponding inset panels of the $x-z$ plane. From left to right, we show: simulation B100R53 in the quiescent phase, B100R53 in the active phase, and B1R53 (where the jet is always active). Each panel also has pink velocity streamlines.}
    \label{fig:drag}
\end{figure*}

\subsection{The influence of magnetic outflows on gas drag}
\label{subsec:results_drag}

As the BH accretes, it will exchange momenta with its gas supply. There are two ways this can happen: 1. through the gravitational forces exerted by the gas onto the BH and 2. through the accretion of linear momentum. We can compare both with the canonical Hoyle-Lyttleton drag rate, which is the flux of linear momentum that enters the accretion cross-section $\pi \ra^2$, 
\begin{equation}
    F_{\rm HL} = \dot{M}_{\rm HL}v_\infty = \frac{4G^2M^2\rho_\infty}{v_\infty^2}
    \label{eq:F_HL}
\end{equation}
This is order-unity with the steady-state gravitational drag force that is derived from linear theory \citep{dokuchaev_1964,ruderman_spiegel_1971,rephaeli_salpeter_1980,ostriker_1999},
\begin{align}
    &F_{\rm SS} = F_{\rm HL} I \\
    &I = \frac{1}{2}{\rm ln}\left(1 - \frac{1}{\machinf^2}\right) + {\rm ln}\left(\frac{r_{\rm max}}{r_{\rm min}}\right)
    \label{eq:drag_analytic}
\end{align}
which depends on the logarithm of the maximum radius due to the long-range nature of the gravitational force. 

To measure the drag rate onto the BH, we use a modified Newtonian formula, which is identical to the Newtonian drag at large distances \citep[][]{cruzosorio_2020},
\begin{equation}
    F_{\rm d,x} = \int \rho\frac{x}{r^3} dV
\end{equation}
Since the motion of the BH with respect to the wind is in the $x$ direction, we only consider the $x$ component of the drag. 
This rate depends on the outer radius of the volume chosen. We calculate the linear momentum by integrating the $x$ momentum flux in the radial direction around a spherical shell at the event horizon of the BH,
\begin{equation}
    F_{\rm p,x} = \int T^{\rm r}_{\rm x}\sqrt{-g}d\theta d\varphi
\end{equation}

Previous works have shown good agreement between the analytic results of \cite{ostriker_1999} and hydrodynamic simulations \citep{macleod_2017, beckman_2018}. However, these studies focused on hydrodynamic accretion, while the presence of magnetized outflows may substantially change drag. The recent work of \cite{li_2020} explored this by injecting various outflows into a BHL accretion flow, finding that depending on the nature and strength of the outflow, drag can be diminished or even reverse sign. 

We study our calculated drag forces ($F_{\rm d,x} + F_{\rm p,x})$ for all simulations in Figure \ref{fig:drag}. First, let us turn our attention to the top-right panel, where we plot the radial profiles of the drag forces for each simulation, averaged over times $>10\,\ra/\vinf$. As $\betainf$ decreases, drag becomes less efficient, and in all cases is below the unmagnetized case (labeled $\betainf=\infty$). Each simulation also exhibits negative drag for some values of the enclosed radius, up to about $r=\ra$ in most cases. When $\betainf=1$, the entire profile changes significantly, and remains negative out to $\gtrsim4\ra$.  Since some of our simulations are highly time-dependent due to the intermittency of the jets, in the top left panel we compare time-averaged radial profiles of the drag force for quiescent and active phases in simulation B100R53. Surprisingly, there is not a significant difference between the two phases; drag is only mildly suppressed when the jet turns on, despite the mass accretion rate dropping by an order of magnitude because of the jet feedback (Fig. \ref{fig:timeSeries}). 

The measured drag rates can be explained by the distribution of the gas around the BH. Since the BH moves in the $-\hat{x}$ direction, the gas accumulated along the $x$ axis is a large contributor to the drag force. In the bottom row of Fig. \ref{fig:drag}, we plot three $x-y$ slices of gas density, along with corresponding $x-z$ slices in the inset panels. From left to right, the first two snapshots shown are from the quiescent and active phases of simulation B100R53. What we can see is that despite the jet clearing out a significant amount of material in the active phase, the $x-y$ slices are quite similar. This explains why the drag forces in both phases are similar; the jet mainly removes material from the polar axis, leaving gas along the $x$ axis only mildly perturbed. If we turn to the third simulation shown, B1R53, we see that the flow morphology is significantly different from the other cases. The bow shock is much broader, and there is a large accumulation of dense gas upstream from the BH. Meanwhile, there is a magnetized low-density cavity downstream of the BH. This distribution naturally causes the drag force to be negative, explaining why the $\betainf=1$ drag force is significantly different than the other simulations. 

This argument is further supported by comparing the $\betainf=10$ drag forces to the $a=0$ drag forces (which also has $\betainf=10$). Outside $\ra$, both have roughly equal drag forces, despite the absence of jets when $a=0$. There is a difference at $r<\ra$, where the $a=0$ simulation is more negative than the other one. We hypothesize that this is because the absolute (rather than dimensionless) BH magnetic flux is higher when $a=0$, due to the higher mass accretion rate (e.g., Fig. \ref{fig:timeSeries}). This only occurs at radii $<\ra$, as this is by definition where the accretion flow and the downstream region are divided.

When $\gamma=4/3$, the drag forces are higher than in any of the other magnetized simulations and are very close to the unmagnetized $\gamma=5/3$ simulation. This is because the gas is more compressible at lower adiabatic indices, leading to higher gas densities in the downstream region, which in turn leads to higher drag forces. 

\section{Magnetic fields in astrophysical environments}
\label{sec:discussion}

So far, we have studied jet formation in BHL accretion using a dimensionless setup. If we want to apply what we have learned from our simulations to real environments, we need to know both the magnetic field strength and geometry in the environment. Since we mainly varied $\betainf$, we will leave the discussion of magnetic field geometry to future work. In Figure \ref{fig:betaContours} we plot characteristic regions in the $\rho-\beta$ plane for three different environments where BHL-like accretion can take place.

\subsection{Interstellar medium}

The multi-phase interstellar medium (ISM) is complex, spanning several orders of magnitude in density and temperature, and features strong magnetic fields \citep{elmegreen_2004}. While it's difficult to assign a general value of $\beta$ to the ISM, there are a wide range of observations \citep{troland_2008,koch_2012,soler_2013} and simulations \citep{li_2009,vazques-semadeni_2011,hill_2012} that we can use to guide our expectations. In this case, we can look to Figure 10 of \cite{kritsuk_2017}, who modeled interstellar magnetized turbulence and found that $\beta$ typically ranged from $\lesssim10$ to $\gtrsim0.1$ across the multi-phase ISM. So, we choose these values as our `typical' range of $\beta$s found in the ISM when outlining it in the $\rho-\beta$ plane in Figure \ref{fig:betaContours} (orange). For $\rho$, we assumed that the ISM ranges from `cold' ($\sim10\,{\rm K}$) to `warm' ($\sim10^4\,{\rm K}$), that in the warm phase the number density is $1\,{\rm cm^{-3}}$, and that the ISM obeys an ideal gas equation of state. Given the low values of $\beta$ in the ISM, our simulations suggest that there is plenty of magnetic flux for a BH to form a jet. Whether or not a jet forms in practice would depend on the magnetic field structure in the ISM. However, due to the low gas densities in the ISM, the corresponding mass accretion rates are also low. This means that the feedback produced by wandering BHs would likely be modest compared to other ISM feedback processes (e.g., supernovae). The jetted emission from BHs wandering through the ISM may provide opportunities to study isolated BH populations in the Milky Way \citep[e.g., ][]{kimura_2021a,kimura_2021b}.

\subsection{The envelopes of red supergiants}

\begin{figure}
    \centering
    \includegraphics[width=\textwidth]{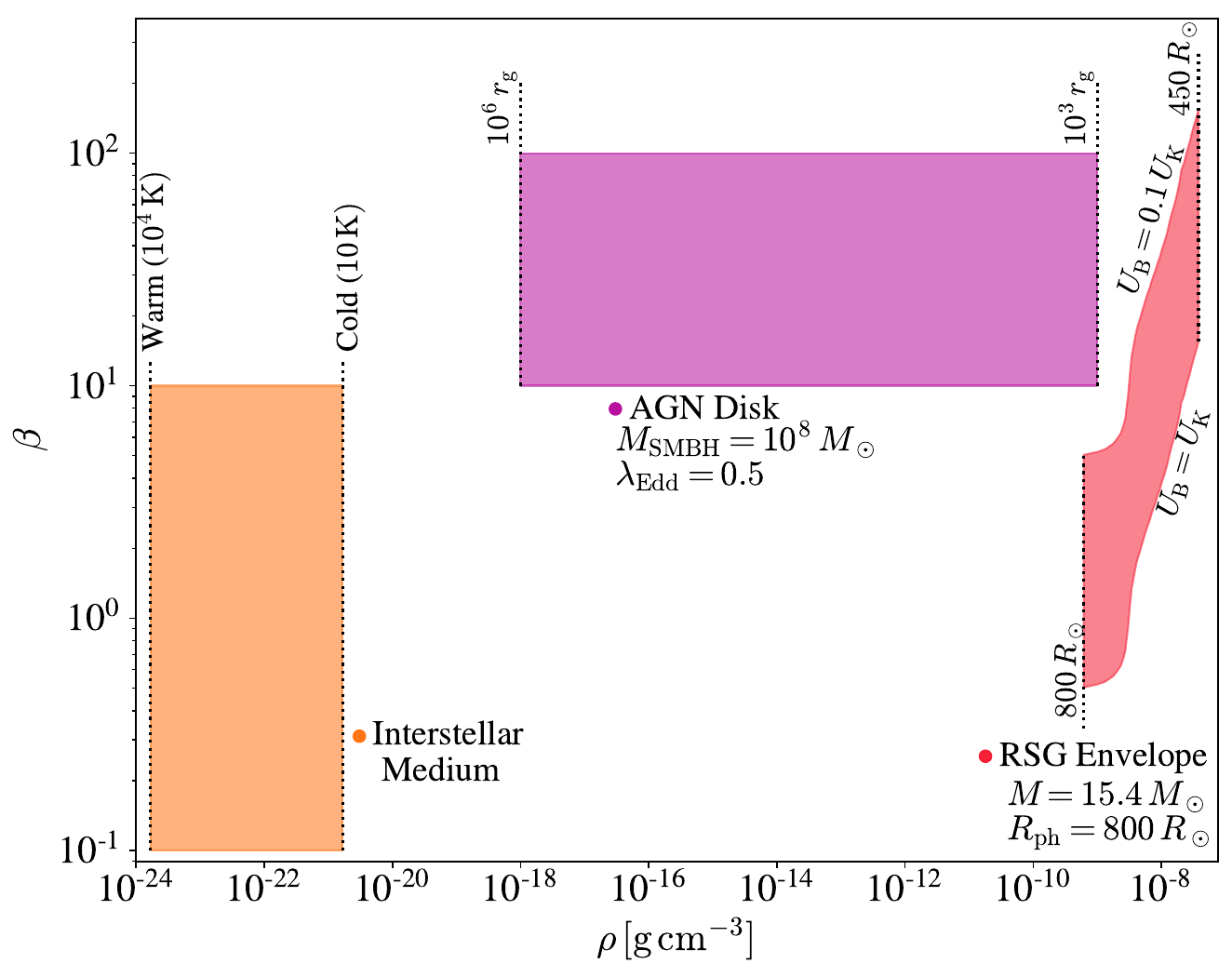}
    \caption{A variety of astrophysical environments have values of $\beta$ that span a wide range of values, $0.1\lesssim\beta\lesssim10^2$, explored in this work. Here, we plot characteristic $\rho-\beta$ planes for three example environments in which BHL accretion onto a BH can be a useful description. These include the interstellar medium (orange), the disks of active galactic nuclei (purple), and the envelopes of red supergiants (red). Here, $\lambda_{\rm Edd}\equiv L/L_{\rm Edd}$ is the Eddington ratio of the AGN and $R_{\rm ph}$ is the radius of the photosphere of the red supergiant. See \secref{sec:discussion} for motivations for each range of values.}
    \label{fig:betaContours}
\end{figure}

In close binary systems, unstable Roche lobe overflow can lead to common envelope evolution. The hydrodynamics of common envelopes is complicated, and simplified descriptions can be useful to gain insight. If the mass ratio of the binary is below $1/3$, then a local wind tunnel approximation can be used to focus on the vicinity of a lower mass companion plunging through the envelope of the more massive star \citep{rosa_2020}. In this case, the hydrodynamics is similar to BHL accretion, distinguished mainly by the local density structure of the envelope and the external gravity of the host star. 

The evolution of a high mass binary system can lead to a black hole becoming embedded in the envelope of a red supergiant (RSG). If the BH forms a jet, it will deposit energy into the envelope, potentially unbinding it and halting the common envelope phase \citep{armitage_livio_2000, mendez_2017}. As studied in this work, the ability of the BH to form a jet is sensitive to the strength of the ambient magnetic field. However, the magnetic field strength and geometry within RSGs is poorly understood. RSGs are often slowly rotating and have high Rossby numbers, making it unlikely that an $\alpha\Omega$ dynamo\footnote{Here, $\alpha$ parameterizes the strength of a magnetic dynamo due to the twisting of field lines, and $\Omega$ is the angular rotation rate of the star.} can produce a large-scale magnetic field \citep{Durney_Latour_1978}. 
Still, observations of RSGs have found that their surface magnetic fields are typically on the order of 1 Gauss \citep{grunhut_2010,auriere_2010,tessore_2017,mathias_2018}. Because RSGs have such extended envelopes, it is unlikely that this surface magnetic field is from a fossil field, but it can be explained by a local convective dynamo operating in the envelope. This is also supported by magnetohydrodynamic simulations of Betelgeuse \citep{dorch_2004}, which found that a convective dynamo can produce typical magnetic field strengths of 100 Gauss in the envelope, which is slightly above equipartition with the kinetic energy of the fluid flows. 

Furthermore, accretion in the common envelope wind tunnel is distinguished from BHL by having non-zero angular momentum, so in principle a dynamo in the accretion flow itself could produce a strong enough poloidal magnetic field to power the BZ mechanism, independent of the background magnetic field strength in the envelope. We emphasize that the assumption of a magnetic dynamo is contentious, and as such our following estimates for the field strength in a convective envelope should be taken as speculative.

For our purposes here, we assume that a convective dynamo operates within the RSG envelope, and that the magnetic energy, $U_{\rm B}$, either reaches equipartition or sub-equipartition levels with the kinetic energy of the convective eddies, $U_{\rm K}$. To determine realistic values of $U_{\rm K}$, we used the results of \cite{goldberg_2021}, who performed hydrodynamic simulations of convection in RSG envelopes. In particular, we used their radial profiles of gas density and turbulent pressure in their 3D RSG2L4.9 model to calculate $\beta$ as a function of $\rho$, assuming $0.1<U_{\rm B}/U_{\rm K}<1$. This is shown in Figure \ref{fig:betaContours}, where we can see that typical $\beta$ values range from about $1$ to about $100$, a range that is covered by our parameter space. This is interesting because it suggests that jets may be able to be produced in BH-RSG common envelopes, contributing to the energetics of the system and potentially producing exotic transients. In future work, we plan on studying the BH common envelope in detail. 

\subsection{Disks of active galactic nuclei}

In the centers of galaxies, the disks of active galactic nuclei (AGN) can interact with the stellar population in the surrounding nuclear cluster. There has been recent interest in the population of stellar-mass BHs (sBHs) that may become embedded in these disks through dynamical capture \citep{tagawa_2020} or via star formation happening within the Toomre-unstable regions of the disk \citep{toomre_1964,stone_2017}. When this happens, the embedded BHs can migrate, accrete at super-Eddington rates, and potentially produce electromagnetic transients. The accretion geometry of the sBH-AGN binary depends mainly on the mass ratio of the system and the AGN disk scale height \citep{kaaz_2021}. If the mass ratio is large, the system resembles more traditional circumbinary disks, where the sBH can impact the structure of the host disk \citep{baruteau_2011,li_2021}. If the mass ratio is low, such as for more massive SMBHs, then the sBH has a more negligible affect on its structure, and the flow can be examined as a local BHL-like wind tunnel. In this case, the main difference from traditional BHL flow is that there is non-zero angular momentum from the shear in the host AGN disk. 

There is a wide array of theoretical support and indirect observational evidence that AGN disks are at least weakly magnetized. Firstly, there is widespread consensus that angular momentum transport in accretion disks is mainly driven by the magnetorotational instability \citep[MRI;][]{balbus_hawley_1991}. The $\alpha$ parameter associated with the efficiency of angular momentum transport is typically inferred from observations to be $\sim0.1-0.4$ \citep{king_2007}, which is consistent with $\beta\lesssim10-100$ in shearing box simulations of MRI-driven dynamos \citep{salvesen_2016}. Recent analytic models have suggested accretion disks are highly magnetized \citep{begelman_2015}, and it has been argued that this is more compatible with observations of accretion disks in both AGN and XRBs \citep{begelman_2017,dexter_2019}. The seed magnetic field in AGN disks (which can be further enhanced by a dynamo) is commonly thought to be provided by galactic magnetic flux threading the gas supply of AGN, which is supported by observations of a large-scale poloidal magnetic field in the Milky Way galactic center \citep{nishiyama_2010}. Finally, the observed presence of relativistic jets in many AGN requires large-scale poloidal magnetic flux to thread the inner parts of AGN disks.

In Figure \ref{fig:betaContours}, we plot the $\rho-\beta$ parameter space associated with AGN disk in purple. We use the Toomre-unstable region (where sBHs are more likely to be embedded) in the models of \cite{sirko_goodman_2003}, which ranges from roughly $10^3$ to $10^6\,r_{\rm g}$ for a $10^8M_\odot$ supermassive BH accreting at half the Eddington rate. We assume that $\beta\lesssim100$, but that the disk midplane is still gas pressure dominated, and so choose $10<\beta<100$. This is enough for embedded sBHs to form jets, and the feedback they exert on their environment can contribute to the pressure support in AGN disks or produce electromagnetic signatures that would signal their presence.

\section{Summary}
We have performed and analyzed the first GRMHD simulations of jet launching by accretion onto rapidly rotating BHs that move through a magnetized gaseous medium. We have adopted a vertical ambient magnetic field and set the accretion radius, $\ra$, to $200\,r_{\rm g}$. In our simulations, we have primarily altered the ambient gas-to-magnetic pressure ratio $\betainf$, which we varied from $1$ to $200$ for a BH spin $a=0.9$ and polytropic index $\gamma=5/3$. Our main findings are,
\begin{itemize}
    \item \textbf{The wind drags magnetic flux towards the BH, which accumulates near the event horizon and causes the flow to become magnetically arrested. } This is clear from Fig. \ref{fig:timeSeries}, where $\phi$ reaches values $\gtrsim50-80$ for $10\leq\betainf\leq200$, which is what is found in simulations of magnetically arrested disks \citep[e.g.,][]{tchekhovskoy_2011}. However, this is \textit{not} true of our $\gamma=4/3$ simulation B10R43, where $\phi$ only reaches $\sim10$. 
    \item \textbf{The significant magnetic flux accumulated in the inner accretion flow allows relativistic jets to form with high enough power to overcome gas pressure and escape the gaseous over-density surrounding the BH} (Fig. \ref{fig:jetStruggle}). After launching, the jets become bent by the incident wind and propagate downstream. These jets can reach extremely high efficiencies of $\eta\sim200-300\%$, despite the gas supply having no net angular momentum. Because the jets push out the infalling gas when they escape, they cause a roughly order of magnitude drop in the mass accretion rate. 
    \item \textbf{The jets vary in activity due to stochastic behavior in the gas and magnetic flux distribution in the inner accretion flow.} At $\betainf\lesssim50$ and $\gamma=5/3$, the jets are essentially on $100\%$ of the time, but at $\betainf=200$ they are only on roughly $10\%$ of the time (Fig. \ref{fig:betaSeries}). The overall jet power decreases with increasing $\betainf$. It is possible that there is a maximum $\betainf$ where jets can be launched, but it is above the range of values that we have adopted here. Jet activity is likely dependent on $\ra/\rg$, which we have set to be $200$ in all simulations. In reality, $\ra/\rg$ is usually much larger, however Eq. \ref{eq:ma_timescale} suggests that larger values of $\ra$ would make it \textit{easier} to produce powerful jets. This is essentially because at larger values of $\ra$ and the same value of $\betainf$, the magnetic flux will be concentrated within a smaller area $\approx \rg^2$.
    \item \textbf{In all simulations with rotating BHs, the jet power saturates to a value at or near the Blandford-Znajek jet power} (Fig. \ref{fig:timeSeries}). The jet power in all but two simulations has an equilibrium value that is very nearly equal to $P_{\rm BZ2}$. The exceptions to this are the jets produced in simulations B200R53 and B10R43. In the former, the supplied magnetic flux is low, and the jet is unable to reach full power before being quenched. In the latter, the jet is highly intermittent and never turns on for long durations, preventing it from reaching full power. Both of these simulations still reach values that are roughly $\sim0.5\,P_{\rm BZ2}$. 
    \item \textbf{The accumulation of magnetic flux in the downstream wake behind the BH redistributes gas and causes drag forces to be much less efficient than in unmagnetized BHL accretion, sometimes becoming negative}. With decreasing $\betainf$, the drag force lowers, and can be roughly an order of magnitude lower than in the hydrodynamic case. This result is relatively insensitive to the presence of jets, as the drag force depends mostly on the distribution of gas along the $x$ axis, while the jet clears out gas near the polar ($z$) axis. Inside the accretion radius ($\lesssim\ra$), the drag force due to the enclosed gas can be negative, which would cause the BH to accelerate rather than slow down. When $\betainf=1$, the drag force is negative for a much wider radial range, up to about $\lesssim5\,\ra$, leading to an even more pronounced effect.
\end{itemize}

Our results have important implications for a variety of astrophysical environments where BHL accretion is a useful approximation. This includes common envelope evolution, where jet feedback could potentially unbind the envelope and the reduced drag efficiency would slow the inspiral. This would increase the post-common envelope semi-major axes of remnant BH binaries, making it less likely for them to merge within a Hubble time. If BHs in binaries accrete from a magnetized stellar wind, then the reduced (or negative) drag forces found here could also alter the angular momentum evolution of the binary. If stellar-mass BHs are embedded in AGN disks, then they could potentially launch jets that escape the host disk. If these jets are observable, they would provide a way of detecting embedded BHs.  

When we consider applying our results to real environments, it is important to recognize that our simulation setup is idealized. For instance, there is no guarantee that the ambient magnetic field will be ordered, as we have assumed here. This would likely decrease the overall efficiency of jet formation. Additionally, for numerical reasons we are forced to set $\ra=200\,r_{\rm g}$ in our simulations. In real systems, $\ra$ can be several orders of magnitude larger, and it is not yet clear how significantly this will affect our results. However, it may very well be that the flow eventually becomes MAD regardless of the size of $\ra$, and in that case the jet efficiency would be unaffected. 

We leave the community with several potential future avenues of study using the GRMHD-BHL framework presented here. This includes varying other parameters, such as the magnetic field geometry, the tilt of the BH, the value of $\ra$ or the Mach number of the wind. Our study can be extended to BHs in common envelopes or wind-fed binaries by including the necessary changes to the structure of the wind and including the external gravity of the companion. Depending on the densities in the modeled system, cooling and radiation physics can also be important considerations, as supported by our $\gamma=4/3$ simulation. This could impact the ability of the BH to accumulate the magnetic flux, launch the jets, and create a magnetized wake, which would affect the drag force.

\begin{acknowledgments}
We thank the anonymous referee for helpful suggestions that improved the paper. We thank E. Ramirez-Ruiz and F. Rasio for useful comments. NK is supported by an NSF Graduate Research Fellowship. AMB is supported by NASA through the NASA Hubble Fellowship grant HST-HF2-51487.001-A awarded by the Space Telescope Science Institute, which is operated by the Association of Universities for Research in Astronomy, Inc., for NASA, under contract NAS5-26555.  K.C. is supported by a Black Hole Initiative Fellowship at Harvard University, which is funded by grants from the Gordon and Betty Moore Foundation, John Templeton Foundation and the Black Hole PIRE program (NSF grant OISE-1743747). The opinions expressed in this publication are those of the authors and do not necessarily reflect the views of the Moore or Templeton Foundations. AT is supported by the National Science Foundation grants AST-2009884, AST-1815304, AST-1911080, OAC-2031997, and AST-2107839. This research was partially carried out using resources from Calcul Quebec (http://www.calculquebec.ca) and Compute Canada (http://www.computecanada.ca) under RAPI xsp-772-ab (PI: Daryl Haggard). This research also used HPC and visualization resources provided by
the Texas Advanced Computing Center (TACC) at The University
of Texas at Austin, which contributed to our results via the LRAC allocation AST20011 (http://www.tacc.utexas.edu).
\end{acknowledgments}

\appendix
\section{Unmagnetized Bondi-Hoyle-Lyttleton Accretion}
\label{app:bhl}

Hydrodynamic BHL accretion has been studied at length in the literature, so we can use previous works to verify that we can accurately simulate unmagnetized BHL accretion. In this appendix, we present a simulation of unmagnetized BHL accretion onto a non-rotating BH (labeled NBNR53 in Table \ref{table:sims}) and compare its flow morphology, mass accretion rate and drag force to previous works. However, most prior simulations of BHL accretion are non-relativistic, but because our accretion radius is much larger than the gravitational radius, we don't expect this to make a substantial difference. 

\begin{figure}
    \centering
    \includegraphics[width=0.8\textwidth]{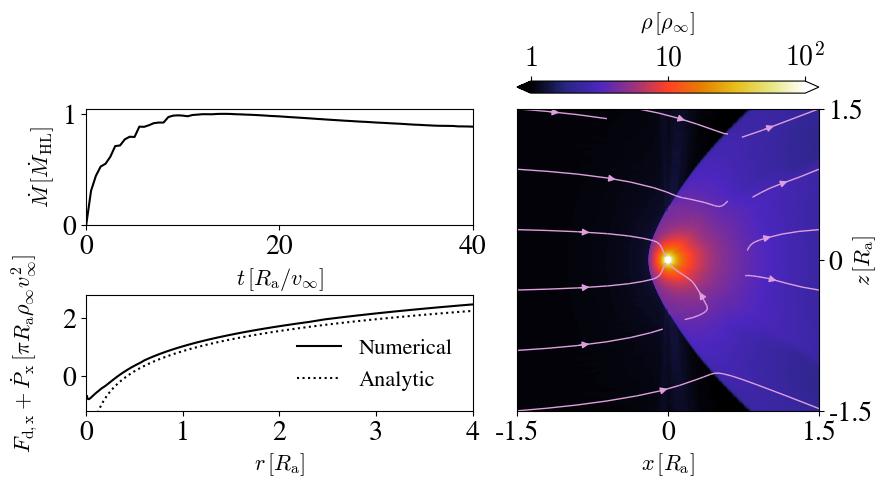}
    \caption{Here, we depict our results for unmagnetized general-relativistic BHL accretion, simulation NBNR53. \textbf{Upper left.} The mass accretion rate is plotted as a function of time, normalized to the Hoyle-Lyttleton accretion rate. The mass accretion rate is in quasi-steady state after $\sim20\,\ra/\vinf$ and is less than but within an order unity factor of $\dot{M}_{\rm HL}$. \textbf{Lower left.} We plot the drag forces (which include gravitational drag and accreted linear momentum) experienced by the BH. The solid curve is our numerical results and the dotted line is the analytic prediction given in Eq. \ref{eq:drag_analytic}, where we set $r_{\rm min}=\ra$. \textbf{Right.} Here we depict a density slice in the $x-z$ plane with velocity streamlines at $t\sim40\,\ra/\vinf$. The standard features of BHL accretion are present: the supersonic wind forms a bow shock, the bow shock is disconnected from the central object, and streamlines are focused behind the BH and are accreted within a dense wake.}
    \label{fig:app_bhl}
\end{figure}

In the right panel of Figure \ref{fig:app_bhl}, we plot the gas density of an $x-z$ slice of our unmagnetized BHL simulation, NBNR53, where $a=0$ and $\gamma=5/3$. Here, we capture the canonical features of BHL accretion: streamlines focus within roughly $\sim\ra$ of the BH, forming a bow shock with some stand-off distance from the central object. Most of the gas is then accreted from the back of the BH \citep{edgar_2004,blondin_2012}. 

In the top left panel of Fig. \ref{fig:app_bhl}, we measure the mass accretion rate of our unmagnetized simulation. We see that the mass accretion rate grows rapidly over the first $10\,\ra/\vinf$, and afterwards gradually tapers to a quasi-steady state at $0.92\,\dot{M}_{\rm HL}$ (averaged at times $>20\,\ra/\vinf$). This is roughly consistent, but slightly higher, than what previous hydrodynamic simulations have found. \cite{blondin_2012}, for instance, found a lower steady state mass accretion rate of $0.56\,\dot{M}_{\rm HL}$ at $\machinf=3$ with a sink radius of $0.01\,\ra$. In hydrodynamic simulations, the accretor is modeled as a `sink' with some radius and an absorbing boundary condition that takes the place of the event horizon. In the higher resolution simulations of \cite{xu_2019}, they found a steady accretion of $0.68\,\dot{M}_{\rm HL}$ at $\machinf=3$ with a sink radius of $0.04\,\ra$. 

We suggest that this discrepancy is due to a difference in the inner boundary condition. It is well known that BHL accretion is sensitive to the choice of inner boundary condition \citep{macleod_2015,xu_2019}. The fundamental reason for why the non-convergence of the mass accretion rate with inner boundary exists is because the sonic surface of the flow is always attached to the accretor \citep{Foglizzo_1997}. In a curved spacetime, the geometry of this sonic surface changes, which can explain the deviation in mass accretion rate. A dedicated numerical study of general-relativistic BHL accretion is warranted to verify this. 

We also include the instantaneous drag profile in the lower left panel of Fig. \ref{fig:app_bhl}. This is the same profile labeled $\betainf=\infty$ in Fig. \ref{fig:drag}. We compare our numerical result (solid line) to the analytic prediction given in Eq. \ref{eq:drag_analytic} (dotted line), where we set $r_{\rm min}=\ra$, and find good agreement.

\section{Raw time series}
\label{app:raw_time_series}

\begin{figure}
    \centering
    \includegraphics[width=0.6\textwidth]{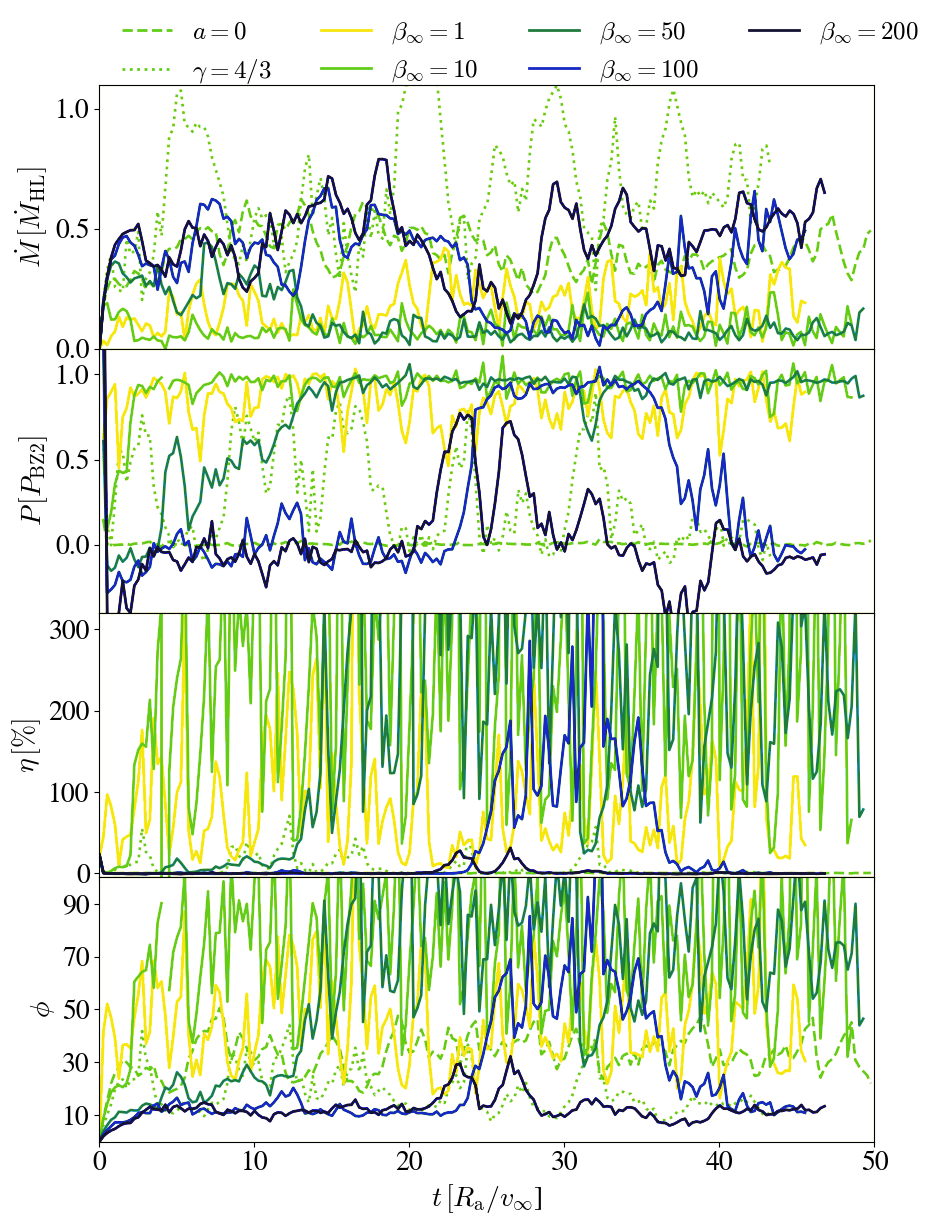}
    \caption{Same as Fig. \ref{fig:timeSeries}, except with no smoothing in time.}
    \label{fig:app_timeSeries}
\end{figure}

In Figure \ref{fig:app_timeSeries}, we plot all of the same time series shown in Fig. \ref{fig:timeSeries}, except without using the moving average in time. We see that $P/P_{\rm BZ2}$ is  similar, whereas all other quantities are much more variable. Our choice to smooth the data in Fig \ref{fig:timeSeries} is done purely for readability. The cadence of outputs plotted here was $0.25\,\ra/\vinf$. 

\section{Dependence of Metric Orientation}
\label{app:tilt}
\begin{figure}
    \centering
    \includegraphics[width=0.8\textwidth]{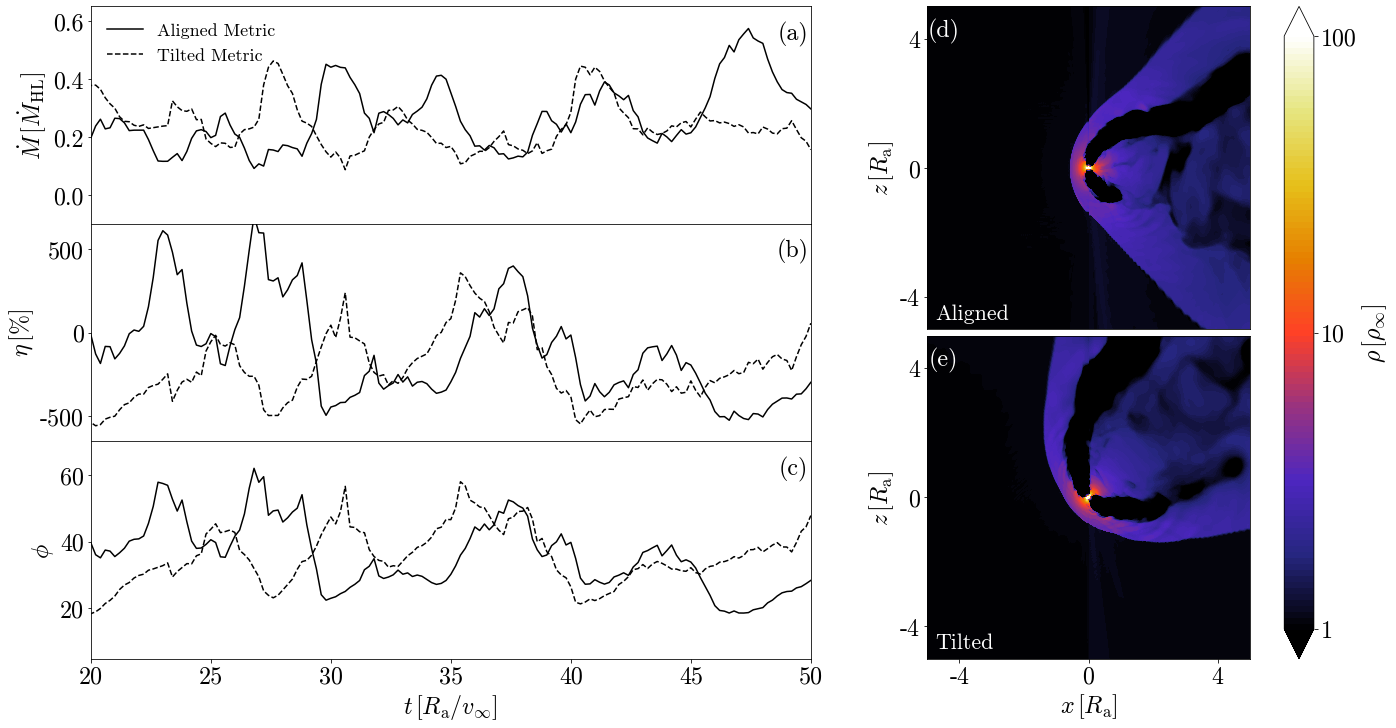}
    \caption{We show results from our two $\ra=50$, $\betainf=100$ simulations where we test the effect of metric orientation on our grid. Our `aligned' simulation adopts our usual configuration, where the BH spin is oriented along the $z$ axis, while in the `tilted' simulation the BH spin and the wind direction is rotated about the $y$ axis by $45^\circ$. \textbf{Panels (a)-(c).} We show the mass accretion rate $\dot{M}$, jet efficiency $\eta$ and dimensionless magnetic flux $\phi$ for both simulations. \textbf{Panels (d)-(e).} We show $x-z$ slices of fluid frame gas density ($\rho$) for each simulation at time $t = 50\,\ra/\vinf$. }
    \label{fig:app_tilt}
\end{figure}

The polar boundary condition of spherical grids has often been challenging for codes to handle, particularly in configurations such as ours where the flow is not axisymmetric about the vertical axis. This is especially true for maintaining the zero-divergence condition of the magnetic field. To show that our polar boundary, which is transmissive \citep{liska_2018,HAMR}, does not affect our results we have performed two comparison simulations with $\ra=50$, $\betainf=100$, $a=0.9$ and $\gamma=5/3$, ran until $t=50\,\ra/\vinf$. In the first we keep our usual configuration where the black hole spin is oriented in the vertical direction and in the second we tilt the metric and the direction of the wind by $45^\circ$ about the $\hat{y}$ axis. We show the results of these simulations in Fig. \ref{fig:app_tilt}. In panels (a), (b) and (c), we show the mass accretion rate $\dot{M}$, jet efficiency $\eta$ and dimensionless magnetic flux $\phi$, respectively. In panels (d) and (e), we show $x-z$ slices of the fluid frame gas density ($\rho$) of both simulations. In general, in all panels, we see that the features are largely the same and we find that in a time-averaged sense they agree. Specifically, averaged over steady-state ($t>20\,\ra/\vinf$) we find that $\dot{M}/\dot{M}_{\rm HL} = (0.269\pm0.11,0.24\pm0.08)$ and $\phi = (35.8\pm10.1,35.8\pm8.4)$, respectively for aligned and tilted runs.

\bibliographystyle{aasjournal}
\bibliography{references}

\end{document}